\documentclass{aa}
\usepackage{graphicx}
\usepackage{txfonts}
\usepackage{natbib}

\begin{document}

\title{A prominent relativistic iron line in the Seyfert 1 MCG-02-14-009 }
\author{D. Porquet\inst{1}
}

\offprints{D. Porquet}
\mail{dporquet@mpe.mpg.de}

\institute{
Max-Plank-Institut f\"{u}r extraterrestrische Physik, Postfach 1312,
D-85741, Garching, Germany
}

\date{Received  / Accepted }

\abstract{
I report the discovery of a prominent broad and asymmetrical feature near
 6.4\,keV in the Seyfert 1 \object{MCG-02-14-009} (z=0.028) with
 {\it XMM-Newton}/EPIC.
 The present short X-ray observation (PN net exposure time $\sim$5\,ks) 
is the first one above 2\,keV for MCG-02-14-009.
 The feature can be explained by either 
a relativistic iron line around either a Schwarzschild (non-rotating)
or a Kerr (rotating) black hole. 
If the feature is a relativistic iron line around a Schwarzschild black 
hole, the line energy is 6.51$^{+0.21}_{-0.12}$\,keV 
with an equivalent width of 631$^{+259}_{-243}$\,eV and that the inclination 
angle of the accretion disc should be less than 43$^{\circ}$.
  A relativistically blurred photoionized disc model gives a very
  good spectral fit over the broad band 0.2--12\,keV energy range. 
The spectrum is reflection dominated and this would indicate that the primary source
in MCG-02-14-009 is located very close to the black hole, where
gravitational light bending effect is important (about 3--4\,R$_{\rm
  g}$), and that the black hole may rapidly rotate. 
\keywords{
galaxies: active -- X-rays: galaxies --
accretion discs -- quasars: individual: MCG-02-14-009}
}
\titlerunning{A prominent relativistic broad line in MCG-02-14-009 }
\authorrunning{D. Porquet}
\maketitle

\section{Introduction}

The Fe\,K${\alpha}$ line complex observed in the 6--7\,keV range is
 an important spectral tool. 
Indeed the determination of the origin of the Fe\,K${\alpha}$ line 
is the most direct signature of the inner accretion disk around black hole (BH).
With the satellite {\sl ASCA} several observations of broad and 
asymmetrical (red wing) lines were discovered with the most famous in 
\object{MCG-06-30-15} (e.g., \citealt{Tanaka95, Fabian02}). 
 These line are commonly interpreted as emission originated 
from the inner part of the accretion disk around the massive BH  
(e.g., \citealt{Tanaka95,Fabian95,Nandra97,RF97}, 
and see also the recent exhaustive reviews from
 \citealt{Fabian00} and \citealt{FM05} for more detailed information). 
The current generation of X-ray satellites {\sl XMM-Newton} and {\sl Chandra} 
 have enlightened that broad line profiles are not so common
 in Active Galactic Nuclei (AGN),  
as previously inferred from {\sl ASCA} data, with the large majority 
of AGN showing the presence of a narrow iron line near 6.4\,keV data 
 (e.g., \citealt{Reeves03,Bianchi04,Page04,P04a,Pi05}).  
From the line characteristics (shape, energy, width, equivalent width ...)
primordial information about the accretion disk 
(inclination, emissivity profile, the inner stable circular orbit) as well as 
 about the BH (spin) itself can be inferred (see e.g., \citealt{FM05}). 

In this paper I present an {\it XMM-Newton} observation of Seyfert 1 
MCG-02-14-009. This is the first observation of this object above 2\,keV. 
 This observation is part of an {\sl XMM-Newton} survey of 21 bright 
{\sl ROSAT} selected AGN with low intrinsic absorption. 
\cite{Gallo05} reports the global properties of this sample.  
 Here I focus on the discovery of a prominent broad and
asymmetrical feature near 6.4\,keV and I investigate the possible
 physical origins of this feature. 

Note that all fit parameters are given in the quasar's rest frame, 
with values of H$_{\rm 0}$=75\,km\,s$^{-1}$\,Mpc$^{-1}$, 
and q$_{\rm 0}$=0.5 assumed throughout. 
The errors quoted correspond to 90$\%$ confidence ranges for one
 interesting parameter ($\Delta \chi^{2}$=2.71).
 Abundances are those of \cite*{An89}.

\section[]{XMM-Newton observation}

MCG-02-14-009 was observed by {\sl XMM-Newton} on September 4, 2000
 (ID\,0103860701; orbit 135; exposure time $\sim$\,9.5\,ks). 
The EPIC MOS cameras \citep{Turner01} were operated in the Small Window mode, 
while the EPIC PN camera \citep{Str01} 
was operated in the standard Full Frame mode. 
The data were re-processed and cleaned (net time exposure $\sim$ 5ks for PN) 
using the {\sc SAS version 6.1.0}
 (Science Analysis Software) package, 
 using the latest {\sc CCF} files concerning the
 PN filter transmission (EPN$\_$FILTERTRANSX$\_$0014.CCF), the effective
 area of the PN X-Ray telescope (XRT3$\_$XAREAEF$\_$0010.CCF), 
and the PN spectral response
 (EPN$\_$QUANTUMEF$\_$0016.CCF, EPN$\_$REDIST$\_$0010.CCF). 
Since pile-up effect was negligible, X-ray events corresponding to patterns
 0--12 and 0--4 events (single and double pixels) were selected for  
MOS and PN, respectively. Only good X-ray events (with FLAG=0) were included.
 A low-energy cutoff was set to 200\,eV. 
The source spectrum and the light curve were extracted 
using a circular region of diameter 30$^{\prime\prime}$ centered on
 the source position. 
Background spectra were taken on the same CCD than the source
 (excluding possible X-ray point sources). 
The {\sc xspec v11.3} software package was used for spectral 
analysis of the background-subtracted spectrum 
using the response matrices and ancillary files derived from the SAS tasks 
{\sc rmfgen} and {\sc arfgen}. 
Since in this work, I focused on the study of the broad feature 
near 6.4\,keV, I used only the PN data which have a much better sensitivity 
above 6\,keV compared to MOS data. However, I checked that  
similar results between the PN and MOS data were obtained. 
The PN spectrum were binned to give a minimum of 20 counts per bin.\\
The signal to noise ratio were not sufficient for reliable RGS data analysis.

\section[]{Spectral analysis}\label{sec:spectra}

In the following, I used the updated cross-sections for X-ray absorption by 
the interstellar medium ({\sc tbabs} in {\sc XSPEC}) from \cite*{Wilms00}.  
In all subsequent fits, I included both the Galactic column density 
($9.31 \times 10^{20}$\,cm$^{-2}$) and a possible 
additional absorption (N$^{\rm int}_{\rm H}$) located at the quasar redshift. 

\begin{figure}[!t]
\resizebox{8cm}{!}{\rotatebox{-90}{\includegraphics{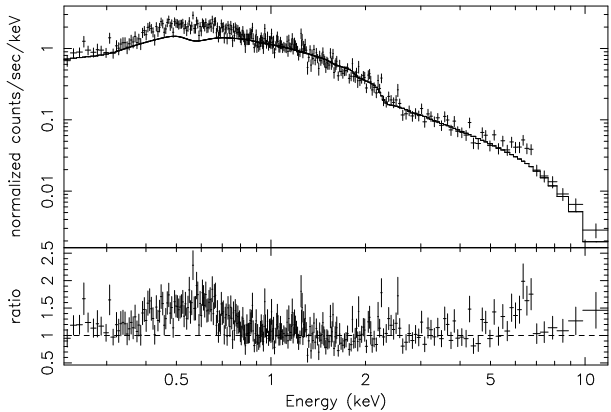}}}
\caption{The PN spectrum of MCG-02-14-009 (in the observer frame).
A power-law has been fitted to the 2--5~keV data
and extrapolated to lower and higher energies.
A soft X-ray positive residual is seen between about 0.4--0.7\,keV, 
as well as a broad and significant positive deviation near 6.4\,keV,
suggesting the presence of a Fe\,K${\alpha}$ line.
For presentation,
the data have been re-binned into groups of 5 bins,
after group of a minimum of 20 counts per bin is used for the fit.
}
\label{fig:spectrum}
\end{figure}

\subsection{The broad-band spectrum}\label{sec:look}

First, a single absorbed power-law model was fitted to 
the broad-band 0.2--12\,keV PN spectrum, but a poor fit was obtained 
 ($\chi^{2}$= 591.2/396) with $\Gamma$=2.12$\pm$0.03,
 N$^{\rm int}_{\rm H}<$0.13 $\times$ 10$^{20}$\,cm$^{-2}$. 
An absorbed broken power-law gave a much better fit of the data 
 ($\chi^{2}$/d.o.f.= 465.7/394) 
with $\Gamma_{\rm soft}$=3.02$^{+0.27}_{-0.24}$, 
$\Gamma_{\rm hard}$=1.87$\pm$0.06, 
E$_{\rm break}$=1.37$^{+0.12}_{-0.09}$\,keV,
 and N$^{\rm int}_{\rm H}$= (8.2$\pm$0.3) $\times$10$^{20}$\,cm$^{-2}$. 
The inferred X-ray luminosities are: 
L(0.2--12\,keV)= 2.9 $\times$ 10$^{43}$\,erg\,s$^{-1}$, 
and L(2--10\,keV)= 6.2 $\times$ 10$^{42}$\,erg\,s$^{-1}$.

To characterize the hard X-ray continuum, an absorbed power-law 
model has been fitted over the 2--5\,keV energy range where the 
spectrum should be relatively unaffected by the presence of a
 broad soft excess, a warm absorber-emitter 
medium, an Fe\,K${\alpha}$ emission line, and a contribution above 8\,keV from 
a high energy Compton reflection hump.
In this energy range, the data were well fitted by a single power-law model
 with  $\Gamma$=1.97$\pm$0.16 ($\chi^{2}$/d.o.f.=92.6/81).  
This power-law slope was similar to the value found in Seyfert type 1
 \citep{Reynolds97,Brandt97}. 
Figure~\ref{fig:spectrum} displays the spectrum 
extrapolated over the 0.2--12\,keV broad band energy. 
A positive residual is seen between about 0.4--0.7\,keV, 
and specially a large and positive asymmetrical residual is clearly seen 
 near 6.4\,keV (in the quasar rest-frame) 
suggesting the presence of a Fe\,K${\alpha}$ line complex.

\begin{table*}[!t]
\caption{Best-fitting spectral parameters of the PN spectrum 
 (exposure time $\sim$ 5\,ks) in the 2--12\,keV energy range
 for an absorbed (Galactic, N$_{\rm H}$=9.31$\times$10$^{20}$\,cm$^{-2}$) 
power-law (PL) component plus a line profile: 
GA: Gaussian profile; and {\sc diskline}  \citep{Fabian89} and {\sc
  laor}  \citep{Laor91}: profile line
emitted by a relativistic accretion disk for a non-rotating BH and 
 a maximally rotating BH, respectively. 
 An emissivity law $q$ equal to -2 was assumed.
 (a): R$_{\rm in}$=6\,R$_{\rm g}$ and R$_{\rm out}$=1\,000\,R$_{\rm g}$.
 (b): R$_{\rm in}$=1.23\,R$_{\rm g}$ and R$_{\rm out}$=400\,R$_{\rm g}$. 
The F-test probability was estimated comparing to the single power-law
model (PL).}
\begin{center}
\begin{tabular}{llllllllll}
\hline
\hline
\noalign {\smallskip}                       
{\small Model}      &  \multicolumn{1}{c}{\small $\Gamma$} &\multicolumn{4}{c}{\small Line parameters}&{\small $\chi^{2}$/d.o.f.} &   F-test   \\
\noalign {\smallskip}                       
                   &                    &  \multicolumn{1}{c}{E (keV)}    &  \multicolumn{1}{c}{$\sigma$ (keV)} &  \multicolumn{1}{c}{$\theta$ (deg)} & \multicolumn{1}{c}{EW (eV)} \\
\noalign {\smallskip}                       
\hline
\noalign {\smallskip}                       
{\small PL }       & {\small 1.83$\pm$0.09}&   \multicolumn{1}{c}{ -- } &   \multicolumn{1}{c}{ -- } &   \multicolumn{1}{c}{ -- } &   \multicolumn{1}{c}{ -- }   &     {\small 145.2/114} & \multicolumn{1}{c}{ -- }\\ 
\noalign {\smallskip}                       
\hline
\noalign {\smallskip}                       
{\small PL + GA}       & {\small 1.92$\pm$0.11} & {\small 6.51$^{+0.16}_{-0.20}$ } & {\small 0.28$^{+0.19}_{-0.17}$} &    \multicolumn{1}{c}{ -- }  &  {\small 527$^{+277}_{-248}$} &  {\small 127.1/111}&   99.80$\%$  \\ 
\noalign {\smallskip}                       
\hline
\noalign {\smallskip}                       
{\small PL + {\sc diskline$^{(a)}$}} & {\small 1.93$\pm0.10$} & {\small 6.51$^{+0.21}_{-0.12}$} &    \multicolumn{1}{c}{ -- } &   $<$43  & 631$^{+259}_{-243}$ & {\small 125.5/111} & 99.90$\%$ \\
\noalign {\smallskip}                       
\hline
\noalign {\smallskip}                       
{\small PL  + {\sc laor}$^{(b)}$} & {\small 1.95$\pm0.11$} & {\small 6.41$^{+0.37}_{-0.18}$} &   \multicolumn{1}{c}{ -- }  &   $<$49  & 765$^{+315}_{-338}$ & {\small 125.4/111} & 99.90$\%$ \\      
\noalign {\smallskip}                       
\hline
\hline
\end{tabular}
\end{center}
\label{tab:line}
\end{table*}

\subsection{The prominent broad feature near 6.4\,keV}

\begin{figure}[!t]
\vspace{-0.8cm}
\resizebox{7.5cm}{!}{\rotatebox{-90}{\includegraphics{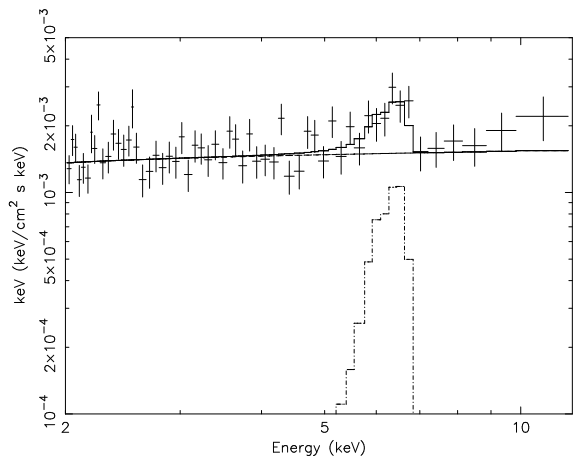}}}
\caption{The unfolded PN spectrum of MCG-02-14-009 (in the observer frame)
 showing the model composed of a power-law continuum and a relativistic
 line profile around a rotating BH ({\sc laor}; \citealt{Laor91}). 
See Table~\ref{tab:line} for the fitting parameter values. 
}
\label{fig:laor}
\end{figure}

As shown in Fig.\ref{fig:spectrum},
a significant deviation is seen near 6.4\,keV in the quasar frame.
Here, I focused on the study of the feature near 6.4\,keV, 
therefore I fitted only the PN spectrum above 2\,keV. 
In the overall 2--12\,keV energy band,
adding a Gaussian line to a single power-law model
drops the $\chi^{2}$ value by 18 with the addition of 3 degrees of freedom
(Table~\ref{tab:line}). This line was significant at 99.8$\%$ according to the F-test.
Adding an ionized emission line and an absorption edge did not improve the fit.
 The line was well resolved with a full width at half maximum  (FWHM) velocity width
of about 30,000\,km\,s$^{-1}$,
 and a large equivalent width (EW) of about 527\,eV (Table~\ref{tab:line}).
The line width and the asymmetrical profile likely indicate that the X-ray emission is
originating from a region close to the BH in MCG-02-14-009.
 The FWHM of the line was only about three times
 smaller than the one found in the Seyfert\,1 MCG-6-30-15,
 which shows the most extreme broad Fe\,K${\alpha}$ line observed up to now
(e.g., 100,000\,km\,s$^{-1}$, \citealt{Tanaka95,Wilms01,Fabian02,Lee02}).
Therefore, I fitted the line with a
profile expected from a relativistic accretion disk around a non-rotating
 (Schwarzschild) BH, using the {\sc diskline} model in {\sc xspec}
from \cite{Fabian89}. I found that such a profile,
 with a inclination less than 43$^{\circ}$ (typical for a Seyfert type 1),
emitted at a rest-frame energy of about 6.5\,keV, provided an excellent 
representation of the line observed in MCG-02-14-009 (Table~\ref{tab:line}). 
The line energy was not enough constrained, according to the error bars,  
to discriminate between a line emitted by ``cold'' iron
 (i.e. $<$\ion{Fe}{xvii}) or by highly ionized iron (i.e. \ion{Fe}{xxv}). 
An equally good fit was obtained (Fig.~\ref{fig:laor}) for a maximally
 rotating BH (Kerr) disc emission line model ({\sc laor}; \citealt{Laor91}).
A higher signal to noise ratio spectrum is required to 
constrain the line energy and to discriminate
between the  Schwarzschild and the Kerr BH, and to determine the BH spin,
 if any.
No narrow ($\sigma$=10\,eV) line at 6.4\,keV, that could be emitted by
 the BLR or the molecular torus, is required by the data.   

\subsection{Ionized Absorption in MCG-02-14-009 ?}

Alternatively, I investigated whether the
strong iron K$\alpha$ feature can be modeled with a warm absorber model. 
Indeed partial covering models where the BH is partially covered by 
an high column density absorbing medium result in a curved continuum that 
could mimic a relativistic line.  
I used the model \textsc{absori} in {\sc xspec} \citep{Done92}.
The model was applied so that the X-ray power-law continuum was modified
 by the absorber (i.e. a partial covering). 
I found a reasonable fit for the continuum ($\chi^{2}/{\rm dof}$=452.2/391)
 but there was still a broad asymmetrical residual feature near 6.4\,keV.
 Adding a relativistic line profile ({\sc diskline},
 assuming an emissivity law $q$=-2), 
 a much better fit ($\chi^{2}/{\rm dof}$=434.0/390) was found 
 with N$^{\rm int}_{\rm H}$=3.5$\pm$1.3$\times$10$^{20}$\,cm$^{-2}$, 
$\Gamma$=2.21$\pm$0.06,
 N$^{\rm absori}_{\rm H}$=(7.5$^{+4.2}_{-2.3}$)$\times$10$^{22}$\,cm$^{-2}$, 
$\xi$=244$^{+107}_{-57}$ erg\,cm\,s$^{-1}$,
 A$_{\rm Fe}$=0.35$^{+0.17}_{-0.13}$ (compared to solar abundance), 
E$_{\rm line}$=6.50$^{+0.22}_{-0.14}$\,keV,
 $\theta_{\rm disk}$=37$^{+14}_{-12}$ degrees, 
and E$W_{\rm line}$=708$^{+264}_{-250}$\,eV.

\begin{figure}[!t]
\vspace*{-0.5cm}
\resizebox{8cm}{!}{\rotatebox{-90}{\includegraphics{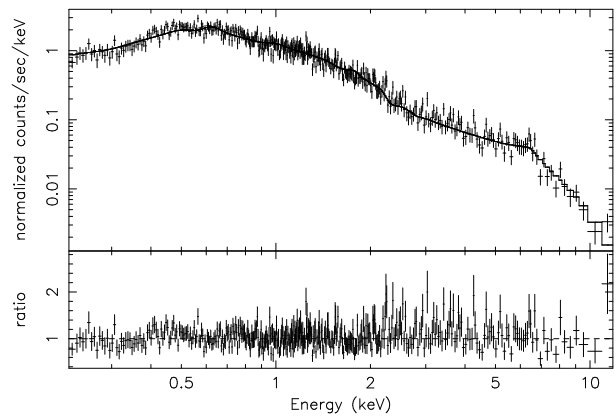}}}
\caption{The PN spectrum of MCG-02-14-009 (in the observer frame).
A relativistically blurred photo-ionized disc model 
({\sc reflion}; \citealt{RF05}) has been fitted, 
and gave a good representation of the whole spectrum. A primary
power-law component is taking into account in the model. 
See text for the fitting parameter values. }
\label{fig:reflec}
\end{figure}

\subsection{The presence of an X-ray ionized disc reflection ?}

 I tested if the whole spectrum is the signature of  
X-ray ionized reflection when the disk surface is irradiated with X-rays 
 using the {\sc reflion} model in {\sc XSPEC} developed by \citealt{RF05}.
Indeed, they have demonstrated that this model can explained in several AGN, 
the broad-band spectrum (0.2--12\,keV) without involving 
several add-oc components such as black-body, power-law continuum. 
 For example, a signature of ionized reflection is the soft excess emission which
 occurs in the 0.2--2\,keV band owing to lines and bremsstrahlung from the
 hot surface layers. It gives a bump in the relativistically blurred spectrum
 which when convolved with an {\sl XMM-Newton} PN response matrix 
mimics a blackbody of temperature 150\,eV. This may explain the 100--200\,eV 
temperature component found in the X-ray spectra of many low-redshift
 PG quasars (e.g., \citealt{GD04,P04a,Pi05}). This model has been used
 successfully for example in the case of the NLS1 \object{PG1404+226} 
 \citep{Crummy05}. \\
The PN spectrum has been fitted taking into account the 
relativistic effect due to relativistic motion in the inner part of
the accretion disc, by blurring the spectrum with a Laor line
profile. A primary power-law ($\Gamma$) was assumed in addition
 to the disc reflection model. 
This model corresponds to the spectrum from a
photoionized disc around a maximally rotating (Kerr) BH. 
I let the inclination of the disc ($\theta$) free and assumed 
an emissivity index of 2 and an inner radius (R$_{\rm in}$) of
 1.23\,R$_{\rm g}$.  As shown in Fig.~\ref{fig:reflec}, 
a very good data fit is found with  ($\chi^{2}$/d.o.f.=418.8/392):
N$_{\rm H}<$ 1.1 $\times$ 10$^{20}$ \,cm$^{-2}$, 
$\Gamma$=1.59$^{+0.11}_{-0.23}$,  
$\xi$=592$^{+133}_{-88}$\,erg\,cm\,s$^{-1}$, 
A$_{\rm Fe}$=0.42$^{+0.59}_{-0.29}$, and 
$\theta<$29.4$^{\circ}$. 
This data fit is much better ($\Delta\chi^{2}$=17.5 for only one
additional free parameter) than the one found in case of a disc
reflection model without taking into account the blurring effect. 
Letting the inner radius as a free parameter, an upper limit 
of about 7\,R$_{\rm g}$ has been found, and therefore is compatible 
with both a non-rotating and a rotating BH.  
However, the flux contribution of the reflection component is about 76$\%$
 compared to the primary power law component, this would mean 
 that in order to produce such
 reflection-dominated spectrum, the primary source
in MCG-02-14-009 must be very close to the BH, where
gravitational light bending effect is important, i.e. about 3--4\,R$_{\rm
  g}$ \citep{MF2004}. Therefore, this would favor a rapidly rotating BH. 

\section{Conclusion}

I report here the analysis of a short {\it XMM-Newton}/EPIC observation 
(PN net time exposure $\sim$ 5\,ks) of the Seyfert\,1 MCG-02-14-009 (z=0.028).
 The present X-ray observation is the first X-ray observation
 above 2\,keV of this object. 
This Seyfert 1 is a weak AGN with a luminosity in the 2--10\,keV 
energy range of 6.2 $\times$ 10$^{42}$\,erg\,s$^{-1}$.

I discovered in the hard X-ray part of the PN spectrum 
a prominent broad and asymmetrical feature near 6.4\,keV. 
I investigated the possible physical origins of this feature
 and I found that it can be explained by either 
a relativistic iron line around either a Schwarzschild (non-rotating)
or a Kerr (rotating) BH. 
A longer X-ray follow-up with a higher signal to noise ratio spectrum
 is required to discriminate between these two possible origins.
If the line is a relativistic iron line around a Schwarzschild BH, 
 the line energy is 6.51$^{+0.21}_{-0.12}$\,keV 
with an equivalent width of 631$^{+259}_{-243}$\,eV and that the inclination 
angle of the accretion disc should be less than 43$^{\circ}$
 (typical for a Seyfert type 1). While a partial covering model
 may explain the underlying continuum, 
it was unable to explain the broad feature.  
Interestingly the equivalent width found here is consistent 
with the one found for average spectrum of the X-ray background sources 
from a large {\sl XMM-Newton} survey of the Lockman Hole field \citep{Streblyanska05}. \\
 A relativistically blurred photoionized disc model gives a very
  good spectral fit over the broad band 0.2--12\,keV energy range. 
The spectrum is reflection dominated and this would indicate that the primary source
in MCG-02-14-009 is located very close to the BH, where
gravitational light bending effect is important (about 3--4\,R$_{\rm
  g}$), and that the BH may rapidly rotate. 

The discovery of this broad and asymmetrical feature 
(probably relativistic iron line) is of great interest since with 
{\sl XMM-Newton} and {\sl Chandra} only few relativistic iron 
line have been confirmed or discovered: 
e.g., MCG-6-30-15 \citep{Wilms01}, \object{MCG-5-23-16} \citep{Dewangan03}, 
\object{Mrk 335} \citep{Gondoin02}, and \object{Q0056-363} \citep{PR03}. 
In conclusion, the study of such relativistic line profiles
 is crucial to understand the effects of strong gravity on the
 accreting material in the vicinity of BH.

\section*{Acknowledgments}

Based on observations obtained with the XMM-Newton, and ESA science
mission with instruments and contributions directly funded by ESA
member states and the USA (NASA). I would like to thank A.\ Fabian and J.\ Crummy for a
helpful discussion about the relativistic blurred photoionized disc
model used in this work. I thank the anonymous referee for fruitful comments.

\end{document}